# Magnetic Force Microscopy Imaging Using Geometrically Constrained Nano-Domain Walls

Héctor Corte-León[1*], Luis Alfredo Rodríguez[2,3], Matteo Pancaldi[4], David Cox[1,5],

Etienne Snoeck[2], Vladimir Antonov[6,7], Paolo Vavassori[4,8], and Olga Kazakova[1]

[1]*National Physical Laboratory, Teddington, TW11 0LW, United Kingdom*

[2]*CEMES, Université de Toulouse, CNRS, 29, rue Jeanne Marvig, B.P. 94347, F-31055 Toulouse Cedex, France*

[3]*Department of Physics, Universidad del Valle, A. A. 25360 Cali, Colombia*

[4]*CIC nanoGUNE, Donostia-San Sebastian, E-20018, Spain*

[5]*Advanced Technology Institute, University of Surrey, Guildford GU2 7XH, United Kingdom*

[6]*Royal Holloway University of London, Egham, TW20 0EX, United Kingdom*

[7]*Skolkovo Institute of Science and Technology, Nobel str. 3, Moscow, 143026, Russia*

[8]*IKERBASQUE, Basque Foundation for Science, Bilbao, 48013, Spain*

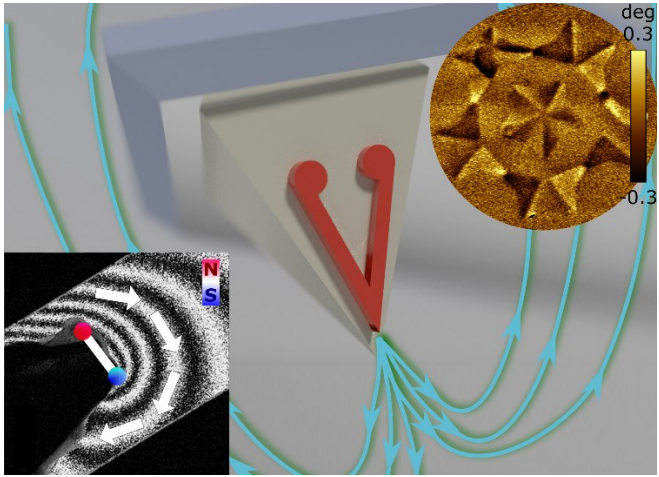

Domain wall probes (DW-probes) were custom-made by modifying standard commercial magnetic force microscopy (MFM) probes using focused ion beam lithography. Excess of magnetic coating from the probes was milled out, leaving a V-shaped nanostructure on one face of the probe apex. Owing to the nanostructure's shape anisotropy, such probe has four possible magnetic states depending on the direction of the magnetization along each arm of the V-shape. Two states of opposite polarity are characterised by the presence of a geometrically constrained DW, pinned at the corner of the V-shape nanostructure. In the other two states, the magnetization curls around the corner with opposite chirality. Electron holography studies, supported by numerical simulations, demonstrate that a strong stray field emanates from the pinned DW, whilst a much weaker stray field is generated by the curling configurations. Using *in situ* MFM, we show that the magnetization states of the DW-probe can be easily controlled by applying an external magnetic field, thereby demonstrating that this type of probe can be used as a switchable tool with a low or high stray field intensity. We demonstrate that DW-probes enable acquiring magnetic images with a negligible interference with the sample magnetization, similar to that of commercial low moment probes, but with a higher magnetic contrast. In addition, the DW-probe in the curl state provides complementary information about the in-plane component of the sample's magnetization, which is not achievable by standard methods and provides additional information about the stray fields, e.g. as when imaging DWs.

Magnetic force microscopy (MFM) enables imaging magnetization distribution with nanoscale resolution, down to ~ 10 nm[1], which is a key requirement for studying novel magnetic nanostructures such as hybrid CMOS structures, e.g. magnetoresistive random-access memory devices[2–4], or magnetic sensors used for Lab-On-a-Chip applications[5,6]. However, smaller and more complex nanostructures require not only higher resolution MFM, but also the ability to extract detailed and accurate information about the magnetization of the sample with minimal invasiveness. Thus, the main challenges that modern MFM technology is facing are i) to increase the sensitivity (both magnetic sensitivity and spatial resolution)[7] and ii) to minimize the mutual interference between the probe and sample magnetization[8–11]. The latter requirement translates naturally into the need to use low magnetic moment probes with high coercivity. However, this recommendation automatically leads to a drastic reduction of magnetic sensitivity, which is presently achievable by employing a high magnetic moment probe.

Different solutions have been previously proposed to tackle the above challenges. For instance, sharp probes with thin magnetic coating provide better spatial resolution[10,12–16], but at the price of a reduced magnetic moment, i.e. a lowered magnetic sensitivity. Thicker coatings, or hard magnetic materials, produce a larger magnetic moment[1,16–20] that improves the magnetic sensitivity and the probe's resilience to magnetic perturbations, but at the cost of a dramatic increase of its interaction with the sample's magnetization and a loss in resolution due to the increased physical radius of the probe's apex.

In general, when using a uniform magnetic coating, its thickness is linked to the sharpness of the probe's apex, as well as the stray field and coercivity of the probe. This does not allow for much freedom in the variation/optimization of the probe[19]. Thereby, the most common scenario is to use a probe matching sample – probe properties in each particular case. However, this approach becomes impractical when the sample under study is magnetically heterogeneous, comprised of areas with high and low anisotropy and magnetization, or with magnetic domain walls (DWs) generating high gradient stray fields[9]. These scenarios are more and more frequent situations faced by modern cutting-edge nanotechnology[21], arising from both the continuous reduction in size and the use of composite, multifunctional materials that are characterized by intrinsically large heterogeneity of properties. In such cases, the stray field generated by the probe may induce changes in the magnetization of low anisotropy regions[22], while regions with high magnetization generate intense stray fields that may affect the magnetic moment of the probe if its anisotropy is not strong enough[23]. Both effects will severely reduce reliability of MFM measurements; hence it would be of the utmost relevance to develop probes with a sufficiently large coercivity, i.e. stability, and with a magnetic moment that can be controllably switched between high and low moment states on demand.

To demonstrate higher spatial resolution, probes modified with magnetic nanoparticles[15,24] or nanowires[8,25–30] were previously demonstrated. By using such magnetic nanostructures, it is possible to increase lateral resolution, but usually at the cost of a reduced magnetic signal and probe stability given by the intrinsically small magnetic moment and anisotropy of the structure.

A different approach utilizes multilayered magnetic probes, designed to have an uncompensated magnetic moment at the probe's apex, which results in the simultaneous increase of probe's magnetic sensitivity and anisotropy[31–33]. However, this approach unavoidably results in a substantial increase of probe radius, with the consequent loss in spatial resolution.

Finally, other approaches involve modification of the MFM technique rather than the magnetic probes, e.g. imaging the same sample twice but with the probe magnetized in opposite directions[34,35]; driving the probe at different frequencies during topography and lift scans[36,37], or using dual probes where one of the probes is non-magnetic and records the topography signal, while the second one (at higher distance from the surface), records the magnetic signal[38]. However, none of the solutions mentioned above fully addresses all the aforementioned challenges.

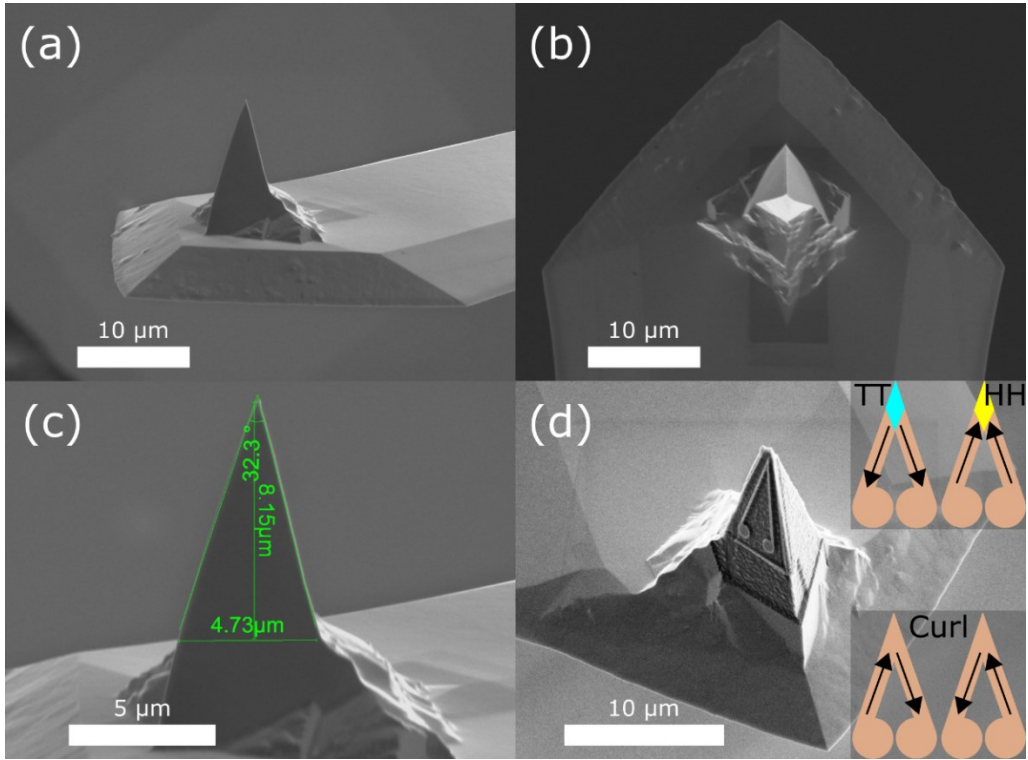

**FIG. 1. SEM images of commercial and custom-made DW-probes. (a) - (c) Commercial NANOSENSORS™ PPP-MFMR AFM probe before ion etching, dimensions of the probe pyramid are shown in (c). (d) V-shape nanostructure obtained by FIB ion etching on one side of the probe apex. Inset: schematics of 4 possible magnetization states with TT/HH DWs marked in turquoise/yellow colour, respectively.**

Here, we propose a novel approach to overcome these problems by implementing geometrically constrained nanosize domain walls (DWs)[39] directly onto MFM probes (DW-probe). This is achieved substituting the conventional uniform magnetic coating with a patterned V-shaped magnetic nanostructure on one side of the probe's pyramid (see Fig. 1). This DW-probe features several characteristic properties that make it suitable for high-resolution imaging of heterogeneous samples. The patterned nanostructure on the probe can be set into four different stable magnetization states [see Fig. 1(d) inset] by applying magnetic field. Two of them are high moment magnetic states with a head-to-head or tail-to-tail (HH or TT, respectively) DWs that have opposite polarity (HH-DW with north polarity and TT-DW with south polarity) at the V-apex of the nanostructure [40,41]. The other two states at the V-apex are low moment "curl" magnetic states caused either by removal or annihilation of a DW. The presence of the disks at the end of the V-shaped nanostructure favors the nucleation of reversed magnetization, thus facilitating the nucleation and annihilation of DWs, i.e., switching the magnetic moment of the tip from low to high via the external magnetic field. At the same time, the geometrical constraint provided by the V-shaped nanostructure bestow an exceptional stability to the 4 states against local magnetic perturbations (e.g., intense stray fields from the sample). Thus, our approach seems to effectively address the aforementioned challenges that need to be met for studies of heterogeneous samples.

We first present a detailed study of the four states of the DW-probe by means of electron holography (EH) investigations[42] and numerical simulations. The remotely controllable switching of magnetization of the DW-probe is then demonstrated by *in situ* MFM imaging. Eventually, the performance of the DW-probe is compared to that of commercial low moment probes by imaging a particularly challenging reference magnetic nanostructure.

**RESULTS AND DISCUSSION**

Figure 2 displays the results of EH experiments, where Figs.2 (a-f) were taken with pyramid aligned along +$z$-axis as shown in Fig. 2(g). Images in Figs. 2 (a-c) show the electron beam phase shift (in radians) due to the stray field emanating from the DW-probe at remanence after applying a pulse of external magnetic field in order to change the magnetization state of the probe (from TT to curl to HH). Images in Figs. 2 (d-f) show the magnetic flux line representations of the corresponding phase shift images in Fig. 2(a-c) (see details in the *Experimental Methods* section).

The different magnetization states obtained through the process described in the *Experimental Methods* are shown in Fig. 2 and correspond to the TT [Figs. 2(a) and (d)] and HH [Figs. 2(c) and (f)] configurations, as well as the curl states [Figs. 2(b) and (e)] (only one curl state is shown). The TT and HH states generate a strong stray field emanating from the apex, which corresponds to a large **monopole-like magnetic charge[39] localized at the outer corner of the V-shaped nanostructure** (i.e. equivalent charge distributions schematically shown as 'north pole' in red for HH and 'south pole' in blue for TT configurations, with size of the circles indicating the dominating pole character). The curl state [Figs. 2(b) and (e)] creates field lines that close around the apex of the probe, which corresponds to a weak **magnetic dipole-like charge aligned perpendicular to the V-structure bisector**. The stray field images shown in Fig. 2 agree with the interpretation of the V-shaped nanostructure acting as a four-state device, where two states have a DW pinned at the apex of the probe with strong emanating stray field and two states without DW and much weaker stray field.

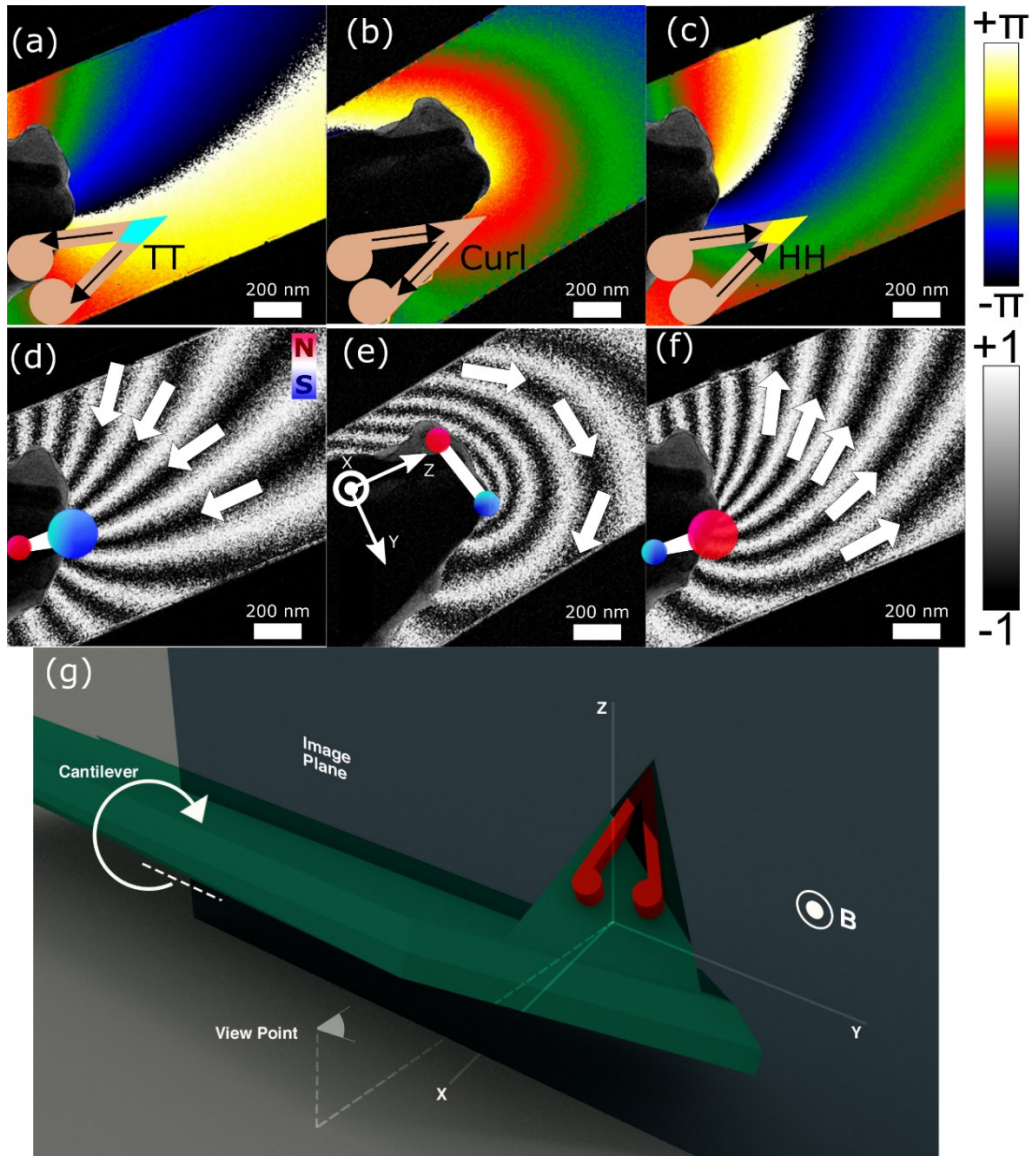

**FIG. 2. EH results for the DW-probe: (a-c) phase and (d-f) magnetic flux images. (a) and (d) TT configuration (i.e. a TT-DW trapped at the corner). (b) and (e) curl state configuration (i.e. no DW at the corner). (c) and (f) HH state configuration (i.e. a HH-DW trapped at the corner). Schematics in (d-f) shows dipole approximation of the stray field, with arrows indicating field direction and size**

**of the circles representing the relative strength of the magnetic poles. (g) Schematics of the relative orientation of the probe and the image plane: the probe's pyramid is oriented along + $z$-axis during imaging [i.e. images (a-f)]; cantilever is rotated as shown by the white arrow to be oriented along the - $x$-axis when the magnetic field is swept to change the magnetization state of the probe. The coordinates in (e) and (g) are the same.**

This picture is further corroborated by the results of the OOMMF micromagnetic simulations of the V-shape in the DW-probes (Fig. 3). Results shown are at zero field after applying a saturating magnetic field either parallel to the V-shape bisector [top of Fig. 3(a), (b) and (c) for the stray field, magnetization and magnetic charge density distribution, respectively], or perpendicular to it [bottom of Fig. 3(a), (b) and (c) for the stray field, magnetization and magnetic charge distribution, respectively]. The state on top of Figs. 3(a-c) corresponds to HH state, with a DW pinned at the corner of the nanostructure and a strong stray field where the spatial distribution is similar to the field shown in Fig. 2(f). In Fig. 3(c) the monopole-like charge distribution (surface magnetic charge in this case) has been drawn schematically as two circles (i.e. red and blue) of different sizes, for the only purpose to illustrate the dominating magnetic pole type corresponding to the surface magnetic charge located at the outer edge of the V-structure (red contour in Figure). The state shown at the bottom of Figs. 3(a-c) corresponds to the curl state, corresponding to a horizontal dipole-like charge distribution (volume magnetic charge in this case) that results in a stray field where the field lines close across the corner similarly to Fig. 2(e). Thus, the OOMMF simulations displayed in Fig. 3 are in good agreement with the experimental EH results (Fig. 2), for all observed stable states and field distributions.

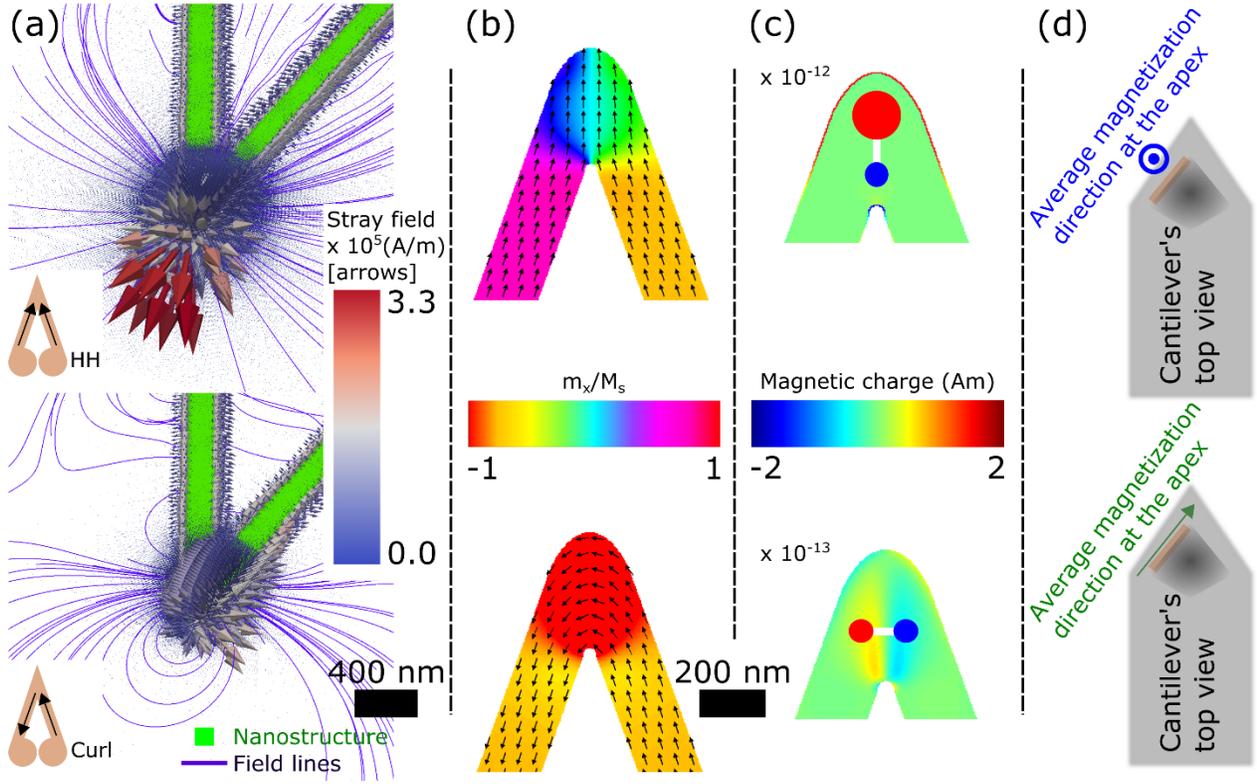

**FIG. 3.** (a) Numerical simulations of the stray field created by the V-shaped magnetic nanostructure when magnetization is either in HH state (top row) or curl state (bottom row). Corresponding images for (b) simulated magnetization, (c) magnetic charge distribution and (d) schematics showing the direction of the stray field during MFM imaging.

We further compared the performance of the DW-probe (resonant frequency $f_0$ = 72.245 kHz, $Q$-factor = 205, and spring constant $k$ = 3.63 N/m) with a commercial low moment probe[43], NT-MDT MFM_LM, with similar mechanical properties ($f_0$ = 65.033 kHz, $Q$ = 195, and $k$ = 2.46 N/m). The two probes were used to scan the same area of a floppy disk [Fig. 4(a)], which is used as a reference sample with a simple magnetization pattern and high coercivity (i.e. suitable to be imaged under applied magnetic fields without perturbing its magnetization). The magnetic imaging was performed in the two-pass standard MFM scan with a lift height of 40 nm, and the oscillation amplitude of both probes was adjusted to be the same (14 nm). The two probes were exposed to the north pole of a permanent magnet prior to the scan. Profiles shown in Fig. 4(a) bottom graph demonstrate that the DW-probe in the TT state shows the greatest phase contrast (~ 0.5 °), whilst the low moment NT-MDT MFM_LM probe has smaller contrast difference (~

0.25 °). This demonstrates that, despite having much less magnetic material than the non-modified probe, the localization of the magnetic moment at the probe apex, enabled by the pinning geometrical confinement of a DW at the corner of the V-shaped nanostructure, results in a sizeable interaction between probe and sample magnetization (via its stray field), two times larger than the one occurring with a low moment probe.

Figures 4 (b) and (c) show the same floppy disk scanned with the DW-probe, while an out-of-plane magnetic field of varying intensity is being applied. In Fig. 4(b) line profiles along directions marked by blue and red dotted lines are shown.

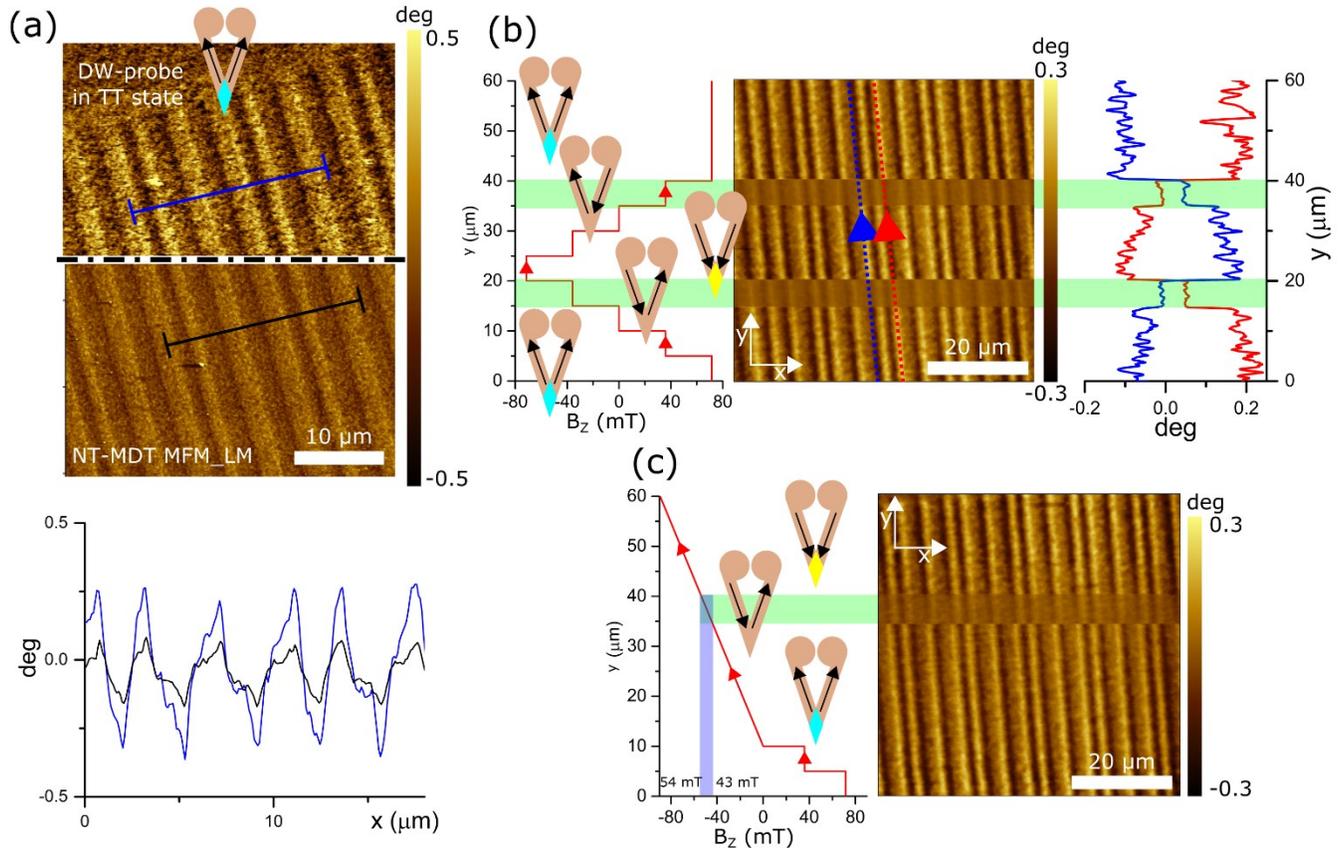

**FIG. 4. (a) Comparison between DW-probe and low moment probe, while performing MFM imaging on the same area of a floppy magnetic disk with the same lift height (40 nm) and oscillation amplitude (14 nm): top – custom-made DW-probe, bottom - low moment NT-MDT MFM_LM probe. Averaged profiles are shown in the graph at the bottom. (b) and (c) *in situ* MFM images taken with a DW-probe in different applied fields. The field along the *z-axis* changes (b) in the step-wise manner to demonstrate different states of the probe and (c) gradually during the scan to extract the probe switching fields. Schematics of the probe states and field evolution are shown on the left**

**hand side in (b) and (c). In all the images, the scan direction *y* is from bottom to top and the frequency of the scan is 0.8 Hz.**

In Fig. 4(b), first, at $y = 0$ μm, a positive magnetic field of 70 mT is applied (i.e. a TT state is induced in the V-shaped nanostructure). At the beginning of the scan, $0 < y < 20$ μm, the field is reduced step-wise to 40, 0, -40, and -70 mT. The image and profiles in Fig. 4(b) demonstrate that for $B = -40$ mT ($15$ μm $< y < 20$ μm) the signal amplitude is reduced, consistently with the nucleation of the curl state in the probe. When the field reaches -70 mT ($20$ μm $< y < 25$ μm), the DW state is opposite to the initial state (i.e. it is now the HH state). This is convincingly confirmed by the switch of the image magnetic contrast as a result of the flip of the DW polarity, see also the plots of the phase signal along the dotted lines marked in Fig. 4 (b). For the rest of the scan ($25$ μm $< y < 60$ μm) the external magnetic field is ramped back up from -70 mT to -40, 0, 40, and 70 mT. This time, the DW-probe is in the HH state at 0 mT ($30$ μm $< y < 35$ μm) and the signal is the opposite of the one observed at $B = 0$ during the ramping down of the field ($10$ μm $< y < 15$ μm). When the applied magnetic field reaches 40 mT ($35$ μm $< y < 40$ μm), the probe is switched into the curl state, producing an MFM signal of much smaller amplitude [Fig. 4(b)]. When the probe reaches top of the scan in Fig. 4(b), the applied magnetic field is $B = 70$ mT, and the state of the probe is the same as it was at the beginning (i.e. a TT state) as seen by restoring the bit contrast back to the initial level in Fig. 4(b).

In order to more precisely identify the DW-probe switching fields, an *in situ* MFM image was taken. First, the probe was saturated with $B = 70$ mT at the beginning of the scan, then the field was reduced to 40 and 0 mT in a step-wise manner, Fig. 4(c). When the probe is scanning between $10 < y < 60$ μm, the field is ramped from zero to negative field values continuously. The MFM contrast in the right-side of Fig. 4(c) shows that the probe transforms into the curl state at -43 mT and eventually switches into HH state at -54 mT. By repeating the cycle several times, (see Fig. S1 in the Supplementary Information for an MFM image with the field ramping up), we can estimate the field required to change a TT or HH state into a curl state being ~40 mT and that for changing a curl state to a TT or HH ~50 mT. This demonstrates that the DW-probe possesses a large coercivity bestowed by the geometrical confinement, i.e., shape anisotropy, which implies a strong stability against external perturbations, including the stray field generated by the sample (it is estimated to be << 1 mT at the probe-sample distance used in the measurement[35,44,45]). Equally relevant is the fact that the shape anisotropy and the resulting coercivity of the magnetic nanostructure are easily tunable by changing the geometrical parameters (width, thickness and length of the arms; the presence/absence of terminating disks, their shape and/or size), without the need to change magnetic material.

To further evaluate the performance of different probes, a patterned magnetic nanostructure made of Py (25 nm thick) was imaged using MFM (Fig. 5). The chosen nanostructure is part of a Penrose pattern used for e-beam alignment in nanofabrication[46,47]. This pattern forms a complex domain structure localized in a small area, enabling comparison of different probes against a particularly complex and challenging case. An approach to compare probes through the real-space probe transfer function and quantitative MFM was recently presented for a similar concept of DW-probes, however, with a different V-shaped design[48].

Topography and MFM images of the Penrose pattern recorded with a commercial low moment NT-MDT MFM_LM probe are shown in Fig. 5(a) and (b), respectively. After taking these images, the probe was replaced with the DW-probe and the next set of images shown in Fig. 5(c) and (d) was taken with the probe in TT and curl states, respectively. Profiles of Fig. 5 (b) to (d) shown in Fig. 5(e) demonstrate that the change in the amplitude is two times larger for the DW-probe (~ 0.6 deg in the HH state) as compared to low moment probe (~ 0.3 deg). Reducing down to ~ 0.25 deg in the curl state. This confirms that the DW-probe in the HH/curl state is characterized by larger/smaller magnetic moments with respect to the commercial probe. Furthermore, the side to side comparison of Figs. 5(b) and 5(c) using the profiles shown in Fig. 5(e) demonstrates that the image recorded with the DW-probe has a higher resolution than that taken with the commercial low moment one, in spite of the similar magnetic contrast (applying the 20-80% Edge Spread Function defined in Standards on Lateral Resolution[49]). The improved resolution of the DW-probe arises from the much higher localization of the magnetic stray field source, the geometrically constrained DW, as compared to the commercial probe. In addition, from comparing Figs. 5(b) to (d) we conclude that the DW-probe interference with sample magnetization is negligible, as it is for the commercial low moment probe. This is further supported by the match between Figs. 5(b) to (d) with the expected magnetic configuration as calculated by micromagnetic simulations, and by the imaging artefacts created by a probe that interacts strongly with the sample's magnetization, as shown in Figs. S2 and S3 in the Supplementary Information.

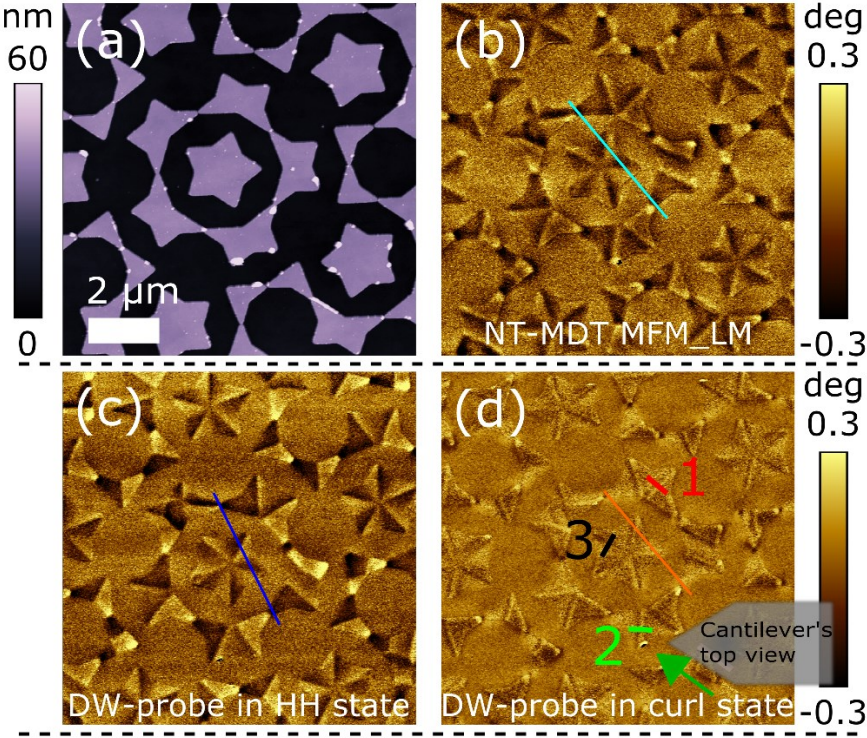

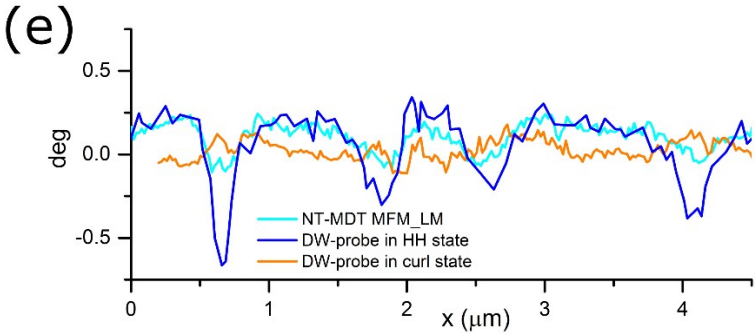

FIG. 5. (a) Topography of a small area of a Penrose pattern. MFM image taken using (b) the low moment NT-MDT MFM_LM probe; the DW-probe in (c) HH and (d) curl states. Red, green and black lines numbered 1, 2, 3 in (d) mark the position of cross-sections analyzed in Fig. 6. Inset: Orientation of the probe during scanning with green arrow indicating the direction of the stray field at the apex. (e) Cross sections taken from turquois, blue, and orange lines in (b) – (f) respectively.

Further inspection of Fig. 5 (b-d) reveals that for all probes characterized by strong stray field (i.e. NT-MDT MFM_LM and DW-probe in HH state) the interaction between domain boundaries in the sample and the probe is a mono-pole like and sensitive to the divergence of the magnetization [i.e. $\nabla(\mathbf{M})$], thus creating a dark-bright sharp transition when the probe scans over a boundary between adjacent domains. This corroborates a distinction between different types of such boundaries, that could occur in samples depending on local anisotropy, exchange interaction, geometrical constraints etc., is generally not achievable with conventional MFM[50]. In this respect, it is remarkable that a closer inspection of Fig. 5(d) shows that the magnetic contrast obtained with the DW/probe in the curl state is **qualitatively** different from the others since it highlights the boundaries in between domains rather than the domains themselves. In Fig. 6 (a)-(c) we further analyze boundary cross sections of Fig. 5(d), see e.g. red, green and black lines. The phase profiles depend on the orientation between the magnetic dipole associated to the DW-probe in the curl state and the domain boundary. This utmost property of our DW-probe in the curl state leads to a sensitivity of the DW-probe in the curl state to the rate of the divergence change [i.e. $\nabla^2(\mathbf{M})$]. This can be appropriately modeled using a simple magnetic dipolar model described in detail in the *Experimental Methods*. Figures 6 (d), (e), and (f) show a schematic of the model, where the probe is represented by a magnetic dipole and the domain boundary by a chain of dipoles representing the charge accumulation created by the DWs in the Penrose pattern. Figure 6(d) denotes the case of probe's dipole in-plane projection parallel to the dipoles in the chain representing the boundary [corresponding to profile 1 in Fig. 5(d)]. The 45° case is shown in Figure 6(e) [see also profile 2 in Fig. 5(d)] and the 90° case – in Fig. 6(f). The 14° angle was introduced to take into account the tilting created by the probe holder. This simple model allows the calculation of the force gradient maps and line plots in Figs. 6(g)-(i). The maps show the force gradient along the $z$-axis ($\frac{\partial F_z}{\partial z}$), when the magnetic dipole of the probe scans over the chain of dipoles describing the boundary. Both the maps and the map profiles taken along the dashed lines demonstrate that the signal measured by the DW-probe in the curl state depends on the relative orientation of the probe and the boundary (whereas the scanning direction is not important). The comparison between the experimental profiles, Figs. 6(a)-(c), and the simulated ones, i.e. bottom panels of Figs. 6(g)-(i), shows an excellent agreement that demonstrates that the DW-probe allows differentiating different types of boundaries in between domains, thereby enabling studies of DWs in planar structures with unprecedented resolution. Moreover, by combining images obtained in the HH/TT states with those taken with the curl state, it is possible to extract 3D information about the sample's magnetization, and hence the DW-probe allows for detailed studies of 3D distribution of magnetization on nanoscale.

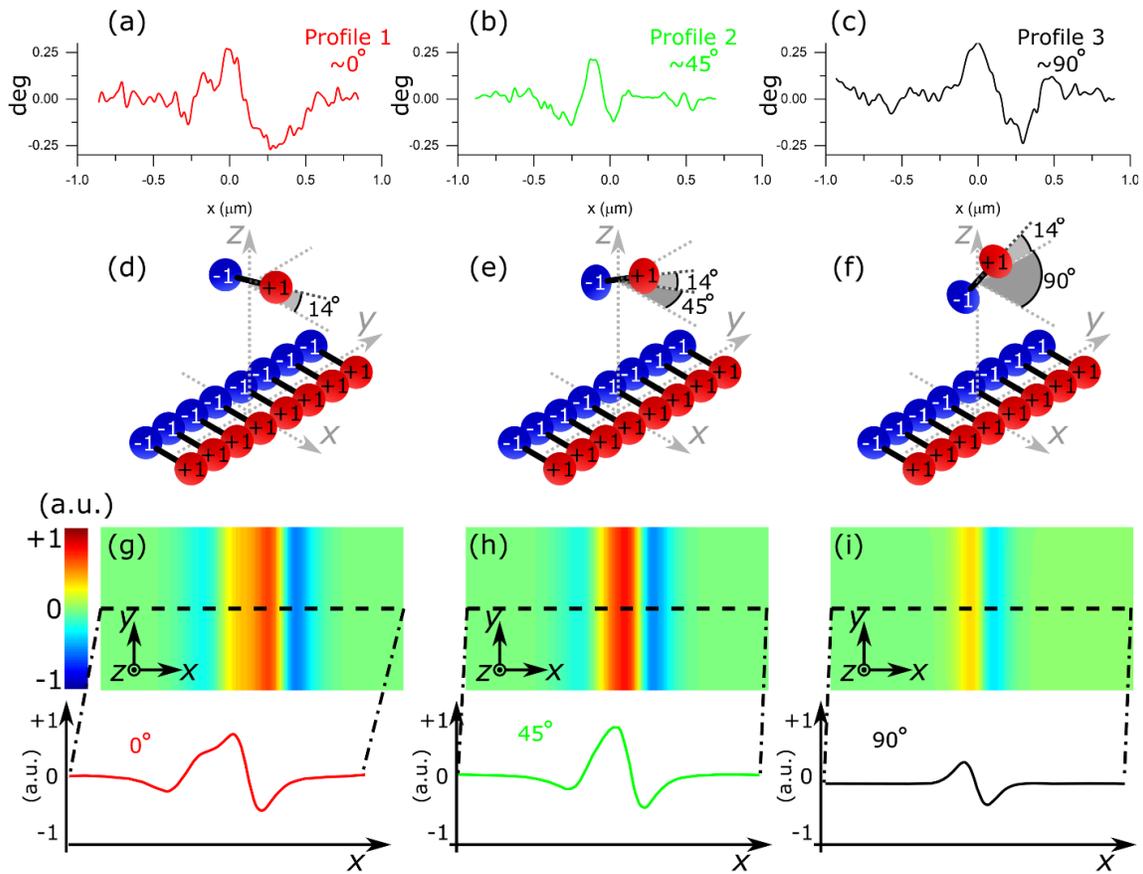

**FIG. 6. Top: MFM phase signal profiles extracted from Fig. 5 (f). Middle: schematic representation of the probe dipole and the surface chain of dipoles. Bottom: calculated $\frac{\partial F_z}{\partial z}$ maps when the dipole representing the probe scans over the chain of dipoles representing the DW. Red, green and black curves in (g), (h) and (i) represent cross sections of the maps shown in (a), (b), and (c) respectively. In all cases, the DW-probe moment is in plane of the sample; the mutual orientation of the probe dipole and the DW are parallel (g); at 45° (h) and perpendicular (i).**

## SUMMARY

We propose and demonstrate a new concept of MFM probes exploiting the peculiar magnetic properties of V-shaped magnetic nanostructures. Such new probes are obtained by FIB-milling of commercial pyramidal MFM probes to leave a V-shaped magnetic nanostructure on one of the sides of the pyramid. This magnetic nanostructure possesses four stable magnetic states, two with a nanometer-size geometrically constrained DW at the corner of the V-shape, and two states without a DW in which magnetization is curling along the corner. An external magnetic field of moderate intensity allows for selecting the desired magnetic state in the probe. EH results and numerical simulations demonstrate that the two DW states (i.e. HH and TT states) generate a strong stray field with opposite polarity in the direction of the probe axis (monopole-like point field sources), while the two states without DW (curl states) produce a much weaker and dipole-like stray field that is perpendicular to the previous one.

*In situ* MFM imaging demonstrated that the phase contrast achieved with the DW-probe in the TT or HH state is twice as large as that from a commercial low moment probe, simultaneously demonstrating **improved spatial resolution**, which arises from the nanometer localization of the magnetic stray field source, i.e., the geometrically constrained DW at the V-shaped nanostructure apex. We also demonstrated that we can controllably set the magnetization of the DW-probe using an externally applied magnetic field, thus enabling different magnetic imaging modes. The large coercivity of the probe, due to the strong shape anisotropy of the V-shaped structure, insures the **stability of magnetic configurations** of the DW-probe that in combination with the low magnetic moment guarantees a **negligible interference** with the magnetic state of the sample. Using a Penrose pattern as a complex and particular challenging test magnetic nanostructure, it was shown that the DW probe is able to achieve better spatial and magnetic resolution and **higher magnetic resolution** with respect to commercial probes of similar magnetic moment. In addition, the DW-probe in curl state is selectively sensitive to the boundaries in between domains, enabling the determination of their internal magnetic structure with unprecedented resolution.

Thus, the DW-probe concept demonstrated here meets the requirements to overcome present main limitations of MFM, namely: combining high sensitivity and spatial resolution with high probe stability and low stray field to minimize interference with sample's magnetization; possibility to adapt probe characteristics to the local sample properties for samples with highly heterogeneous physical properties; sensitivity to fine details of the 3D magnetization distribution at the nanoscale, which are not achievable with conventional MFM. Moreover, the special characteristics of the DW-probe can potentially be used in new MFM imaging modes, e.g. by exploiting stable low and high moment states, it is possible to use these probes in differential phase imaging (switching DW polarity in a double pass approach) or controlled magnetization MFM[35] (four pass approach using the four magnetization states to fully cancel electrostatic interactions). Also, the possibility to set the DW probe in a low moment curl state could, in principle, allow for recording a "true" sample topography by switching off magnetic interactions (setting the probe to a curl state) in the first pass, before measuring the magnetic contrast scanning over the surface at constant height from (setting the tip to a DW state). This is advantageous to the use of conventional magnetic probes, where magnetic interactions are active even during the first topography pass, thus always leading to intermixing of topographic and magnetic signals that limits the quantitative use of MFM[51].

**Experimental methods**

The modified probes were custom-made using magnetically coated commercial probes from NANOSENSORS™ PPP-MFMR AFM[52] and focused ion beam (FIB) to etch away part of the magnetic coating. As it can be seen in Figs. 1 (a) to (c), the probe of these probes has 4 triangular sides which are uniformly coated with a CoCr alloy. Dimensions of the probe's side, when approximated by a triangle, are 8.15 μm in height by 4.73 μm length of the base [Fig. 1(c)]. FIB milling lithography (Ga-ions) is used to etch away the magnetic material from the probe's sides: completely from three sides, while leaving only a V-shaped magnetic nanostructure on one of them [Fig. 1(d)]. The V-shaped nanostructure's arms are 4.48 μm in length by 200 nm in width, and they meet at 32.3º. The estimated thickness[17] of the magnetic coating is about 30 nm. Both arms of the V-shaped nanostructure end in a circular disc of 1μm in diameter to reduce the stray field produced by these parts of the nanostructure[40,53].

Although the exact composition of the magnetic coating of the commercial probes from NANOSENSORS™ PPP-MFMR AFM is unknown, it is expected that it possess in-plane magnetization, and hence the V-shaped structure is expected to behave similarly to ferromagnetic structures of similar dimensions with in-plane magnetization (e.g. 25 nm thick Py L-shaped nanostructures[40,41]), which shows four stable magnetic states depending on the magnetization along the arms.

Micromagnetic numerical simulations were carried out using OOMMF micromagnetic solver from NIST[54]. As shape anisotropy is expected to dominate over magnetocrystalline anisotropy, the cell size used was $5 \times 5 \times 5$ nm$^3$, and the material parameters used were the standard for Py ($M_s = 800 \times 10^3$ A/m, $A = 13 \times 10^{-12}$ J/m, k = 0). Magnetic charge density maps were calculated using $\rho_{Vol} = -\vec{\nabla}\vec{M}$.

The SPM system (Aura from NT-MDT with custom made magnet for applying out-of-plane fields) was used for *in situ* MFM studies in ambient atmosphere and at room temperature.

EH imaging[55] of the stray magnetic field produced by the modified probes was carried out in a Hitachi HF3300 (I2TEM-Toulouse) microscope, a transmission electron microscope (TEM) specially designed to perform *in situ* EH experiments with a high resolution and phase shift sensitivity thanks to a very high brightness cold field emission gun and a spherical aberration corrector (aplanator B-COR from CEOS) to correct the off-axial aberration in both TEM and Lorentz modes[55]. For this study, EH experiments were performed in a corrected Lorentz mode for the normal stage of the microscope. In this stage the sample is placed within the pole pieces of the objective lens which is switched off to favor a free-field condition. However, a controlled magnetic field can be applied by exciting the objective lens, which acts parallel to the electron trajectory. This *in situ* magnetic field capability of the microscope was used to change the magnetic states of the MFM probes. EH holograms were recorded operating the microscope at 300 kV, and using a double bi-prism setup to avoid the formation of Fresnel fringes in the lateral edges of the hologram. The reconstruction of the stray field around the probe was carried out by retrieving, from the holograms, the magnetic component of the phase shift of the object electron wave, $\varphi_{MAG}(z, y)$, which is directly proportional to the magnetic flux[55], $\Phi(z, y)$, [$\varphi_{MAG}(z, y) = (e/\hbar)\Phi(z, y)$, where $e$ and $\hbar$ are the electron charge and the reduced Planck constant, respectively] so images of the phase shift will directly provide maps of the magnetic flux. In addition, $\varphi_{MAG}(z, y)$ and the projected magnetic induction, $B_{proj}(z, y)$ are related as[56] $\nabla\varphi_{MAG}(z, y) \cdot B_{proj}(z, y) = 0$, meaning that the variation of the magnetic phase shift is perpendicular to the direction of the projected magnetic induction, following the right-hand rule between $\nabla\varphi_{MAG}(z, y)$, $B_{proj}(z, y)$ and the electron trajectory. This relationship allow us determining the direction of the magnetic flux.

In order to identify all the stable magnetization states when performing EH, the DW-probe, which is placed inside of the TEM with the pyramid along the + $z$-axis to perform EH [schematically represented in Fig. 2 (g)], is rotated to place the pyramid along the + $x$-axis, and then a strong saturation field is applied towards – $x$-axis [Fig. 2(g)]. According to the probe position inside the TEM, such condition should induce a TT DW. At zero-field condition, the DW probe is tilted back with the pyramid apex pointing in the direction of the + $z$ and – $x$-axes [i.e. at 45º in respect to both $x$ and $z$-axes as depicted in Fig. 2 (g)]. In such configuration, the applied magnetic field was progressively increased until each representative magnetic state was obtained. However, whenever a change in the magnetization was detected, the magnetic field was reduced to zero and the probe was tilted with the apex pointing towards + $z$-axis in order to image the state at remanence.

To facilitate the interpretation of the stray field configuration, magnetic flux line representation [Figs. 2 (d-f)] was produced by applying a sinusoidal function to the amplified magnetic phase shift images [Figs. 2 (a-c)], i.e. *Φ(y, z) ~ cos(nφ$_{MAG}$(y, z))*, where Φ(y, z) is the magnetic flux representation, *n* is an enhancement factor, and *φ$_{MAG}$(y, z)* is the electron beam phase shift.

Simulations of dipolar interaction between the DW-probe in the curl state (approximated as a dipole) and the DWs in the Penrose pattern (approximated as a line of dipoles) were carried out using a simplistic magnetic dipolar model. The curl state was approximated by two magnetic charges $q_1^{probe} = +1$ and $q_2^{probe} = -1$ separated by 1 a.u., and aligned along *x*-axis or *y*-axis [as shown in Fig. 6 (d) and (e) respectively]. A small tilt of 14° was introduced to take into account the tilting created by the probe holder. The DWs in the Penrose pattern were approximated by 200 +1 and -1 surface charges along the *y*-axis as illustrated in Figs. 6 (d) and (e). Each pair of charges is separated by 0.00125 a.u. and the 200 pairs span along 10 a.u. along the *y*-axis. The dipole representing the probe is supposed to scan at a light height of 1.125 a.u. The force used in the dummy magnetic dipolar model was:

$$\vec{F} = \sum_{i=1}^{2} \sum_{j=1}^{200} \frac{q_i^{probe} q_j^{sample}}{r_{ij}^2} \hat{r}_{ij} \qquad (1)$$

**ASSOCIATED CONTENT**

Supporting information:

There are two figures included as supplementary information. The first one is an MFM image, similar to Fig. 4(c) but with the magnetic field sweeping orientation from negative to positive values. The second image is an OOMMF micromagnetic simulation of a portion of the Penrose pattern in one of the stable states at zero field.

The following files are available free of charge.

Magnetic Force Microscopy Imaging Using Geometrically Constrained Nano-Domain Walls – Supplementary Information (PDF).

## AUTHOR INFORMATION

**Corresponding Author**

*E-mail: hector.corte@npl.co.uk

**Author Contributions**

H.C. performed MFM experiments, micromagnetic simulations of the Penrose pattern, and data analysis and wrote the preliminary manuscript. L.A.R. and E.S. performed EH imaging. D.C. fabricated the probes using FIB lithography. M.P. and P.V. performed the micromagnetic simulations of the probe. H.C., V.A., P.V. and O.K. contributed with the general experiment conception and design of the probes. All authors contributed writing the final manuscript.

**Notes**

The authors declare no competing financial interest.

## ACKNOWLEDGMENTS

This work has been partially funded by EMRP and EMRP participating countries under EMPIR project 15SIB06 – Nanomag: Nano-scale traceable magnetic field measurements. This work was also supported by the UK government's Department for Business, Energy and Industrial Strategy, and by the European Union Seventh Framework Program under a contract for an Integrated Infrastructure Initiative Reference No. 312483-ESTEEM2. The authors acknowledge the French National Research Agency under the "Investissement d'Avenir" program reference No. ANR-10-EQPX-38-01" and the "Conseil Regional Midi-Pyrénées" and the European FEDER for financial support within the CPER program". M. P. and P.V. acknowledge support from the Basque Government (program PI_2015_1_19) and from the Spanish Ministry of Economy and Competitiveness through project FIS2015-64519-R (MINECO/FEDER), the Maria de Maeztu Units of Excellence Programme - MDM-2016-0618, and (M.P.) grant BES-2013-063690. The authors are grateful to Patryk Krzysteczko and Hans W. Schumacher for providing the Penrose pattern nanostructure and Robb Puttock for useful discussions.